\documentclass[journal=jctc,manuscript=article]{achemso}

\usepackage[version=3]{mhchem} 

\usepackage{float}

\usepackage{physics}
\usepackage{braket}
\usepackage{mhchem}
\usepackage{color}

\usepackage{tabularx}
\usepackage{siunitx}
\usepackage{xcolor}
\usepackage{hyperref}

\author{Leon Otis}
\affiliation{Department of Chemistry, University of Chicago, Chicago, Illinois 60637, United States}

\author{Yu Jin}
\affiliation{Pritzker School of Molecular Engineering, University of Chicago, Chicago, Illinois 60637, United States}

\author{Victor Wen-zhe Yu}
\affiliation{Materials Science Division, Argonne National Laboratory, Lemont, Illinois 60439, United States}

\author{Siyuan Chen}
\affiliation{Pritzker School of Molecular Engineering, University of Chicago, Chicago, Illinois 60637, United States}

\author{Laura Gagliardi}
\affiliation{Pritzker School of Molecular Engineering, University of Chicago, Chicago, Illinois 60637, United States}
\email{lgagliardi@uchicago.edu}
\alsoaffiliation{Department of Chemistry, University of Chicago, Chicago, Illinois 60637, United States}

\author{Giulia Galli}
\alsoaffiliation{Pritzker School of Molecular Engineering, University of Chicago, Chicago, Illinois 60637, United States}
\alsoaffiliation{Materials Science Division, Argonne National Laboratory, Lemont, Illinois 60439, United States}
\email{gagalli@uchicago.edu}
\affiliation{Department of Chemistry, University of Chicago, Chicago, Illinois 60637, United States}

\title{Strongly correlated states of transition metal spin defects: the case of an iron impurity in aluminum nitride}

\begin{document}

\begin{tocentry}

\includegraphics[width=8.25cm,height=4.45cm]{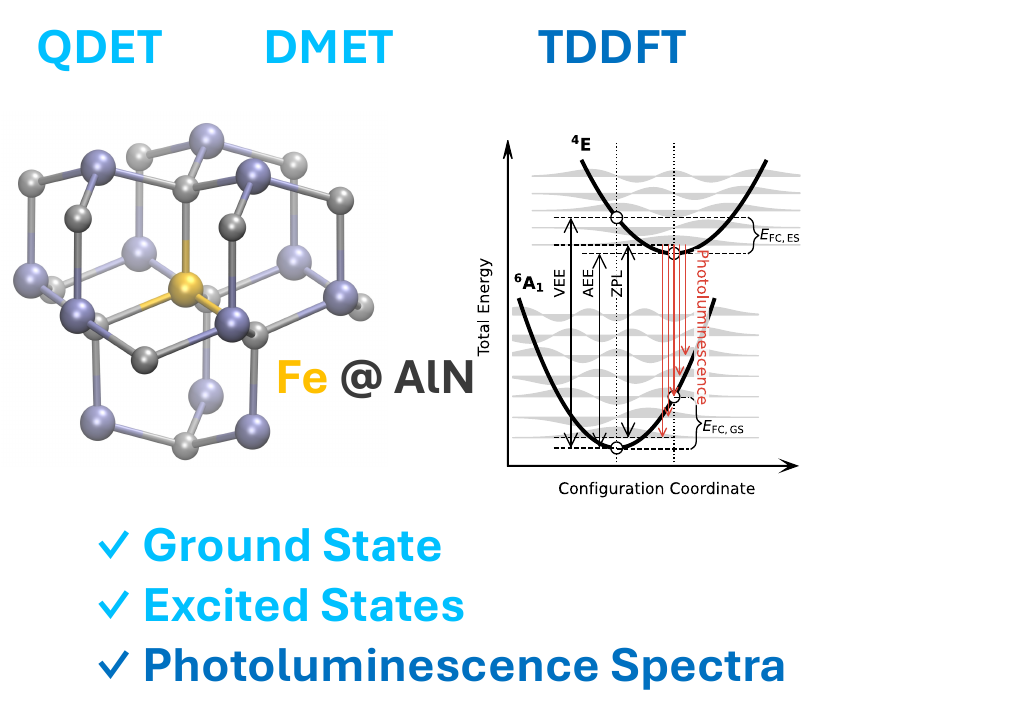}

\end{tocentry}

\begin{abstract}
We investigate the electronic properties of an exemplar transition metal impurity in an insulator, with the goal of accurately describing strongly correlated, defect states. We consider iron in aluminum nitride, a material of interest for hybrid quantum technologies, and we carry out calculations with quantum embedding methods -- density matrix embedding theory (DMET) and quantum defect embedding theory (QDET) and with spin-flip time-dependent density functional theory (TDDFT). We show that both DMET and QDET accurately describe the ground state and low-lying excited states of the defect, and that TDDFT yields photoluminescence spectra in agreement with experiments. In addition, we provide a detailed discussion of the convergence of our results as a function of the active space used in the embedding methods, thus defining a protocol to obtain converged data, directly comparable with experiments.
\end{abstract}


Transition metal impurities in semiconductors and insulators constitute an interesting
category of point defects proposed as qubit candidates \cite{Varley2016} for quantum technologies.\cite{Wolfowicz2021}
In addition, they play an important role
in carrier recombination processes, influencing the performance of materials used for light-emitting diodes and other
electronic devices.\cite{Wickramaratne2019}

First-principles calculations play a key role in characterizing the properties of point defects
and identifying the most promising systems for experimental studies and applications.
Most of the theoretical efforts have so far employed density functional
theory (DFT),\cite{Freysoldt2014,Dreyer2018} although in recent years several papers have appeared using higher levels of theory\cite{Ma2020,Sheng2022,Mitra2021,Haldar2023,Mitra2023,Verma2023,Jin2023} to describe strongly correlated defect states. In particular,
quantum embedding techniques\cite{Vorwerk2022} offer a promising
approach to the study of point defects by treating the localized defect states
with high-level electronic structure methods, while employing a more approximate description for the bulk solid hosting the defect.
Multiple strategies for developing quantum embedding theories have been pursued
in the physics and chemistry communities and various techniques differ in their choice of approximations.\cite{Sun2016}
While different embedding methods have been successfully applied in recent years\cite{Mitra2021,Ma2021,Sheng2022,Muechler2022,Haldar2023,Verma2023,Vorwerk2023} to
several defect systems, particularly in diamond, silicon carbide and magnesium oxide, their performance on transition metal impurities in semiconductors and insulators remains less explored.

\begin{figure}

\includegraphics[width=\columnwidth]{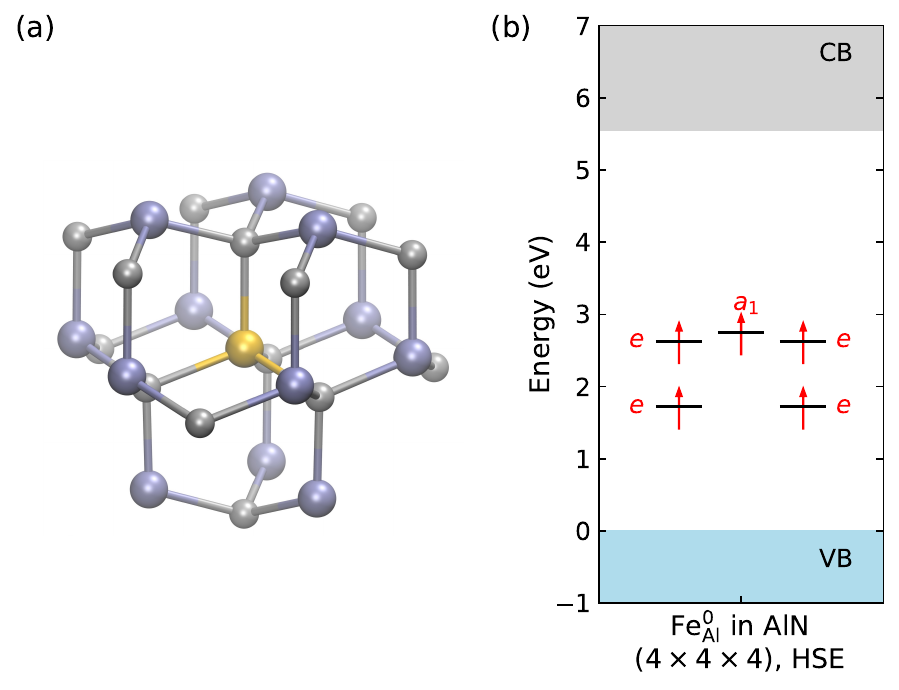}
\caption{(a) Depiction of the Fe (yellow sphere) impurity in AlN. Al and N atoms are represented as purple and gray spheres, respectively. (b) Single-particle defect levels computed at the DFT/HSE level in a $(4 \times 4 \times 4)$ supercell containing 256 atoms, using spin-restricted calculations. The five defect levels are singly occupied, representing the high spin $S = \frac{5}{2}$ ground state of symmetry $\ce{^{6}A_1}$.}
\label{fig:fe_aln_structure}

\end{figure}

Here we focus on a transition metal impurity in an insulator, iron substituting aluminum ($\text{Fe}_{\text{Al}}$) in wurtzite aluminum nitride (AlN) and we compare the results of quantum embedding techniques and time-dependent density functional theory (TDDFT) in describing the electronic properties of the system.
AlN possesses multiple desirable traits as a host material of spin defects,\cite{Weber2010} including a wide
band gap\cite{Li2003} and piezoelectric properties\cite{Fei2018} that may
offer strain-based control schemes\cite{Falk2014,Seo2017,Sohn2018} of qubit states.
In our case study, we consider the neutral charge state of the defect, with Fe in the formal $3+$ oxidation
state and five electrons in its $3d$ orbitals; this state of Fe in AlN  was previously found to be the lowest in energy .\cite{Wickramaratne2019}
The defect possesses $C_{3v}$ symmetry with the Fe-N bond
in the crystallographic $c$ direction being slightly longer than the other three, as determined by DFT geometry optimizations.\cite{Wickramaratne2019,Muechler2022}
Localized single-particle states involving the five orbitals of $e$ and $a_1$ symmetry in the band gap are shown in Figure \ref{fig:fe_aln_structure}.
Crystal field theory
provides a qualitative description of the nature of the many-body states,
with the $\ket{\ce{^{4}T_{1}}}$ and $\ket{\ce{^{4}T_{2}}}$ states of
tetrahedral symmetry splitting into $\ket{\ce{^{4}E}}$ and $\ket{\ce{^{4}A_{2}}}$
or $\ket{\ce{^{4}A_{1}}}$ states with $C_{3v}$ symmetry.
Experimentally, the defect is known\cite{Malguth2008} to have a high spin $S = \frac{5}{2}$ ground
state of symmetry $\ce{^{6}A_1}$ and photoluminescence
spectroscopy measurements have been reported\cite{Baur1994}

Most theoretical treatments of defects in
AlN,\cite{VandeWalle2004,Tu2013,Varley2016,Seo2016,Seo2017} including the Fe
impurity,\cite{Zakrzewski2016,Wickramaratne2019} have relied on
DFT and the application of higher-level electronic structure methods to this system is at an early stage.
A recent study\cite{Muechler2022} applied an embedding method,
based on the constrained random phase approximation
(cRPA),\cite{Aryasetiawan2004,Bockstedte2018} to a set of defects including the
Fe impurity in AlN.
The $\text{Fe}_{\text{Al}}$ defect was found to be a particularly difficult case
for cRPA due to the sensitivity of the predictions to
the choice of the DFT exchange-correlation
functional (PBE or HSE) and the choice of double counting corrections for the embedding method.
Notably, only the combination of PBE without any double counting correction yielded the
expected high spin $\ce{^{6}A_1}$ ground state. However, the agreement with experiments may be fortuitous since double counting corrections are required for a correct application of the theory.

Motivated by the interest in transition metal impurities as possible spin defects and by the theoretical challenges associated with their electronic structure, we study
the $\text{Fe}_{\text{Al}}$ defect in AlN and compare the results of two embedding methods, density matrix embedding theory (DMET)\cite{Knizia2012,Knizia2013,Wouters2016} and quantum defect embedding theory (QDET)\cite{Ma2020,Ma2021,Sheng2022}, with proper double counting corrections.
We demonstrate that the embedding methods succeed in correctly and robustly predicting the character of the defect ground
state and low-lying excited states.
In addition, we make use of spin-flip TDDFT\cite{Jin2023} to obtain excited-state geometries, adiabatic excitation energies, and photoluminescence spectra for comparison with experiments.

We start by summarizing the electronic structure methods
applied in our study.
QDET is a Green's function-based embedding method that constructs the
following effective Hamiltonian for a chosen active space of orbitals

\begin{equation}
\label{eq:qdet_h}
H^{\text{eff}} = \sum_{ij} t_{ij}^{\text{eff}} a_i^{\dag} a_j + \frac{1}{2} \sum_{ijkl} v_{ijkl}^{\text{eff}} a_i^{\dag} a_j^{\dag} a_l a_k
\end{equation}
where the $t_{ij}^{\text{eff}}$ and $v_{ijkl}^{\text{eff}}$ are one- and two-body terms.
QDET uses $G_0W_0$ starting from DFT single-particle orbitals as the low-level electronic structure method to obtain the
effective Hamiltonian parameters and $H^{\text{eff}}$ is solved with full configuration interaction (FCI) within the chosen orbital space.
This formulation enables the use of an exact double counting correction,\cite{Sheng2022,Chen2024} thus avoiding the ambiguities faced by cRPA in choosing among different approximate double counting schemes.\cite{Muechler2022}

The choice of the active space within which $H^{\text{eff}}$ is diagonalized requires special care in quantum embedding theories.
In a previous application\cite{Sheng2022} of QDET to the nitrogen-vacancy and group-IV vacancy defects in diamond, a localization factor $L_n = \int_{v \subseteq \Omega} | \psi_n^{\textrm{KS}}(\textbf{r}) |^2 d\textbf{r}$ was introduced to quantify the degree of localization of the Kohn-Sham wave functions $\psi_n^{\textrm{KS}}(\textbf{r})$ around the defect, where the integration volume $v$ is centered around the defect of interest and is smaller than the supercell volume $\Omega$.

Since a larger value of $L_n$ indicates stronger localization of the wave function around the defect, an active space can be constructed by including all wave functions whose $L_n$ exceeds a prescribed threshold. However, while the most localized defect levels have the largest $L_n$ values, other wave functions associated with the host may exhibit similar values of $L_n$. Consequently, constructing the active space solely based on $L_n$ becomes highly sensitive to the choice (e.g., shape and size) of the integration volume $v$. In this work, we adopt a protocol\cite{Ma2020b,Ma2021,Chen2024} that defines the active space based on the single-particle energies of the wave functions, calculated at the DFT level of theory. Starting from a minimal active space formed by the five defect orbitals (shown in Figure \ref{fig:fe_aln_structure}), we expand it by adding the valence bands of AlN, beginning with the valence band maximum (VBM) and progressing downward in energy. This approach is physically motivated, as wave functions with energies closer to the band gap are more likely to contribute to low-lying excitations.

In contrast to QDET, DMET relies on wave function-based methods with Hartree-Fock as the low-level electronic structure method, avoiding the need for
a double counting correction.
All orbitals from an initial Hartree-Fock calculation on the entire system are transformed into maximally localized Wannier functions (MLWFs)\cite{Marzari1997,Marzari2012}
and can be divided into sets of fragment and environment MLWFs based on the
distance in real space from a defined impurity cluster of atoms.
A Schmidt decomposition of the Hartree-Fock wave function 
\begin{equation}
    \label{dmet_schmidt}
    \ket{\Phi} = (\sum_i \lambda_i \ket{f_i} \otimes \ket{b_i}) \otimes \ket{\textrm{core}}
\end{equation}
is used to
generate a space of fragment and entangled bath orbitals $\ket{f_i}$ and $\ket{b_i}$ respectively; the  Hamiltonian projected into this basis can be treated with a high-level electronic structure method.
In this work we use a periodic version of DMET, beginning with a restricted open-shell Hartree-Fock (ROHF) calculation and employing multireference
methods, complete active space self-consistent field (CASSCF)\cite{Roos1980,Siegbahn1980,Siegbahn1981} and second-order
N-electron valence state perturbation theory (NEVPT2),\cite{Angeli2001a,Angeli2001b,Angeli2002,Angeli2004} as the high-level solvers in the impurity space.\cite{Pham2020}
CASSCF provides a means for describing the potentially strongly correlated nature of defect states while NEVPT2 can account for some of the dynamic correlation neglected by CASSCF.
Recent applications of this incarnation of DMET to the nitrogen-vacancy center in diamond and
the oxygen vacancy in magnesium oxide have made use of extrapolations to
the non-embedding limit to obtain accurate excitation energies.\cite{Mitra2021,Haldar2023,Mitra2023,Verma2023}
The theoretical frameworks of DMET and QDET are rather different and each method approaches the problem from distinctly different directions in terms of
the choice of electronic structure methods; it is thus interesting to compare their performance and understand their respective strengths and weaknesses.

The embedding methods used here can provide excitation energies, but since QDET, CAS-DMET, and NEVPT2-DMET gradients are not available, geometries cannot be optimized at this level of theory for ground and excited states. To compute excited-state geometries \cite{Hutter2003,Jin2023} and spectra, we employ spin-flip TDDFT, where excited states are described as linear combinations of
singly excited configurations generated by a spin-flip excitation operator
from a high spin reference.\cite{Shao2003,Wang2004,Li2012,Bernard2012,Casanova2020}
This formalism has been efficiently implemented in the WEST code\cite{Govoni2015,Yu2022} and
employed to obtain excited-state energies and geometries of defects in diamond\cite{Jin2023}, silicon carbide\cite{Jin2023},
and magnesium oxide\cite{Jin2023}, as well as vibrationally resolved absorption and photoluminescence spectra.\cite{Jin2022}


We carried out calculations in supercells with 108, 192, and 256 atoms and the $\Gamma$ point only, using
experimental lattice constants of $a=3.110\ \text{\r{A}}$ and $c=4.980\ \text{\r{A}}$.\cite{Schulz1977}
Atomic positions were optimized in the presence of the defect at the PBE level using Quantum ESPRESSO\cite{Giannozzi2009,Giannozzi2017,Giannozzi2020}.
QDET and TDDFT calculations were performed with the WEST code\cite{Govoni2015,Yu2022,Jin2023} using a 60 Ry
cutoff for the plane wave basis and SG15 norm-conserving pseudopotentials.\cite{Schlipf2015} All TDDFT calculations adopted the Tamm-Dancoff approximation.
For QDET and TDDFT, we obtained results with the semilocal Perdew-Burke-Ernzerhof (PBE) \cite{Perdew1996} functional, the Heyd-Scuseria-Ernzerhof (HSE) \cite{Heyd2003} hybrid functional, and the dielectric dependent hybrid (DDH)\cite{Skone2014} functional. The DDH functional includes a fraction of exact exchange determined by the inverse of the macroscopic dielectric constant of AlN (0.24), which is similar to the value of 0.25 used by HSE, and we shall see below that both hybrid functionals yield similar results.
For DMET calculations we used the pDMET code\cite{PDMET_mitra} interfaced with PySCF.\cite{Sun2018,Sun2020}
The Wannierization step of DMET was performed with Wannier90\cite{Pizzi2020} and the pyWannier90 interface.\cite{Pywannier_pham}
The impurity cluster chosen in DMET is composed of the Fe atom and the four neighboring N atoms and it is used to define the embedding space of fragment and bath orbitals.
In DMET calculations, we used GTH pseudopotentials\cite{Goedecker1996} and for the two larger supercells we used
a gth-dzvp basis set\cite{VandeVondele2007} on the Fe atom and nearest neighbor N
atoms with a gth-dzv basis on the remaining atoms. For the 108-atom supercell, we used a gth-dzvp basis set on all atoms.
Our DMET results have been extrapolated to the non-embedding limit; however we have found that for this defect computed vertical excitation energies are quite insensitive to the size of the chosen embedding
subspace (see Supporting Information).

We have found that DMET and QDET both correctly predict the ground state of the defect, $\ce{^{6}A_1}$; in the following
we describe the technical parameters required to obtain convergence of the energies with the two methods for the lowest $\ce{^{4}E}$ excited state.

\begin{figure}

\includegraphics[width=0.5\textwidth]{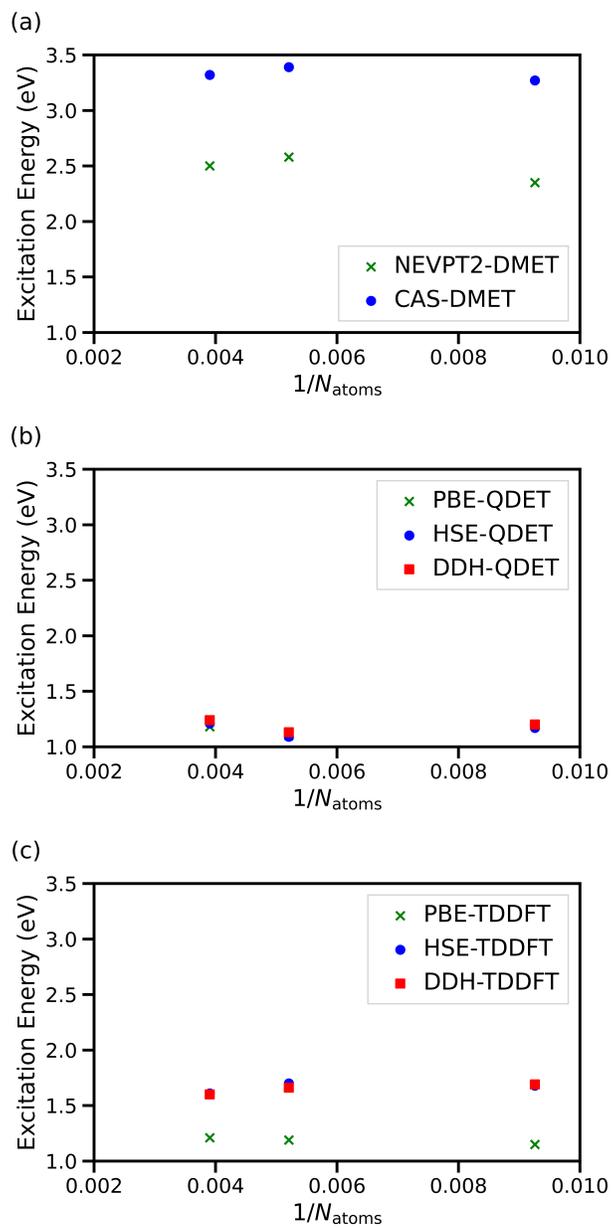}
\caption{Vertical excitation energies of the lowest $\ce{^{4}E}$ state obtained with (a) DMET, (b) QDET, and (c) TDDFT, using 256, 192, and 108 atom supercells.}

\label{fig:vee_4e_supercell}

\end{figure}

We first compare DMET and QDET results obtained with the minimal (5e,5o) active space of the Fe $3d$ orbitals.
For these and all subsequent results that focus only on the ground and doubly degenerate $\ce{^{4}E}$ state, DMET calculations employ a state average over the three states.
As depicted in Figure \ref{fig:vee_4e_supercell}, we observe only limited
changes in the value of the excitation energy with system size for all methods.
The excitation energies obtained with DMET, shown in panel (a), vary by less than 0.2 eV as a function of the supercell size, and NEVPT2-DMET results are consistently lower than CAS-DMET results by 0.8 eV.
QDET's excitation energies, shown in panel (b), exhibit an even smaller variation of at most 0.1 eV as a function of the supercell size and the predictions of QDET are less sensitive to the functional choice, compared to TDDFT.
Excitation energies obtained with TDDFT, shown in panel (c), vary by at most 0.1 eV as a function of the supercell size.
Note that with the minimal active space, DMET excitation energies are significantly larger than those obtained with QDET and
TDDFT, by more than an eV, even when using NEVPT2.

We now turn to exploring larger active spaces, keeping in mind that their size is limited by the use of high-level methods.
The problem of active space selection is an open challenge that has been
extensively studied\cite{Veryazov2011,Stein2016,Sayfutyarova2017,Bao2019,King2021} in molecular calculations.
Our objective here is simply to evaluate the convergence of the lowest $\ce{^{4}E}$ state energy as a function of the active space size.

\begin{figure}

\includegraphics[width=0.5\textwidth]{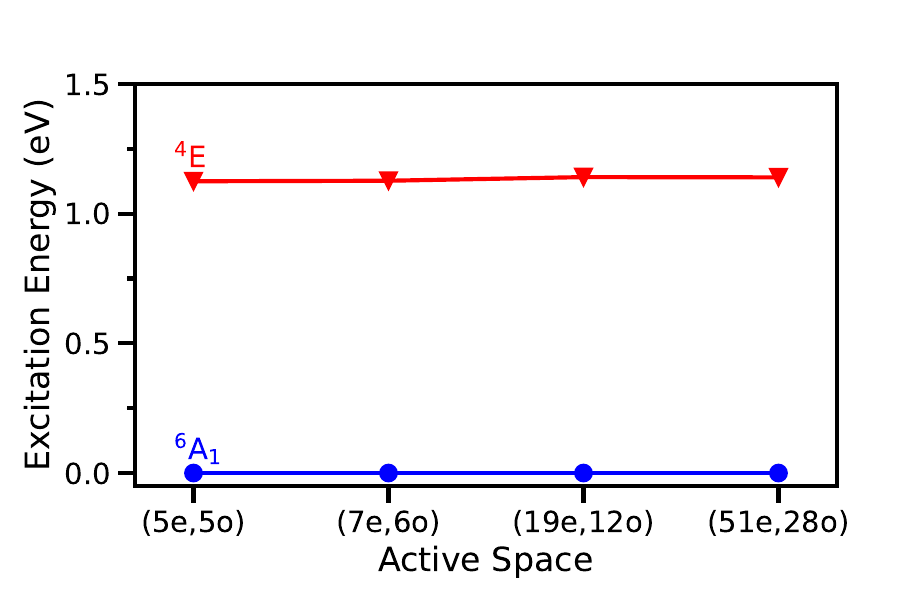}
\caption{Vertical excitation energies of the lowest $\ce{^{4}E}$ excited
state obtained with QDET, in a 256-atom supercell, for different active space sizes, starting from DFT calculations at the PBE level of theory. Single particle orbitals are added to the active
space defined by the localized defect states in order of decreasing energy from the valence band maximum. The (7e,6o), (19e,12o), and (51e,28o) spaces correspond to taking orbitals within 0.2, 0.4, and 0.6 eV of the valence band maximum, respectively.}
\label{fig:as_qdet}

\end{figure}

We begin by discussing the QDET active space using the 256-atom supercell model. Starting with the (5e,5o) minimal active space, we progressively include valence bands of AlN, beginning from the VBM and moving downward in energy. Figure \ref{fig:as_qdet} shows the VEE of the $\ce{^{4}E}$ excited state calculated using QDET with the PBE functional. Results are presented for the minimal active space as well as three expanded active spaces that incorporate valence bands up to 0.2, 0.4, and 0.6 eV below the VBM, respectively. The expanded active spaces remain tractable by FCI due to the limited number of new configurations obtained by adding only occupied valence bands.  We observe that the inclusion of the additional valence bands has only a minor impact on QDET's predicted excitation energy for the lowest $\ce{^{4}E}$ state. 
Similar trends are observed when using the DDH and HSE hybrid functionals, as shown in Table \ref{tab:full_aee_data}. These results demonstrate that QDET can achieve accurate excitation energy predictions for the Fe defect using only the physically motivated (5e,5o) minimal active space, augmented by states close to the VBM.

\begin{figure}

\includegraphics[width=0.5\textwidth]{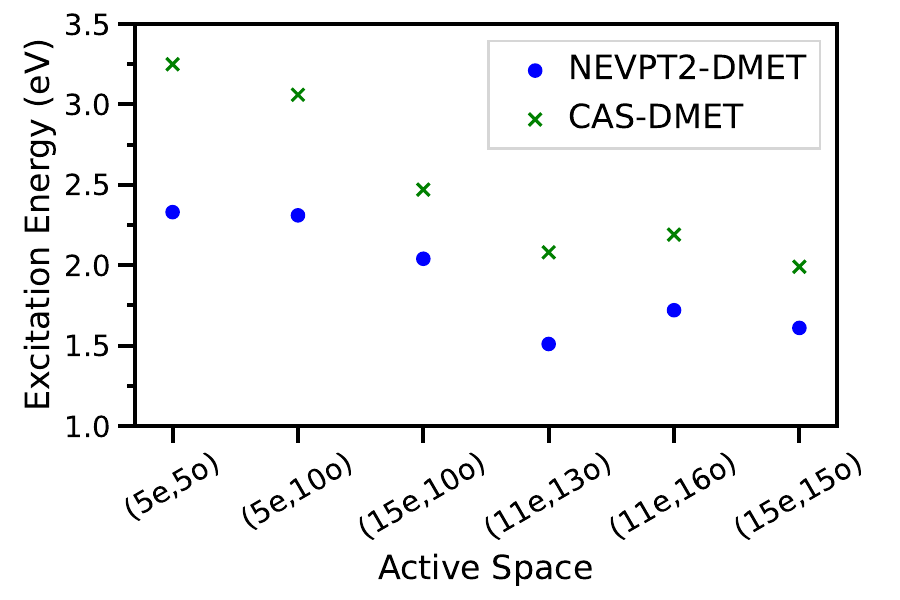}
\caption{Vertical excitation energies of the lowest $\ce{^{4}E}$ excited
state as obtained with CAS-DMET and NEVPT2-DMET in a 108-atom supercell, for different active space sizes.}
\label{fig:larger_as_dmet}

\end{figure}

Next we investigated the active space selection for DMET using the 108-atom supercell model.
Unlike in QDET, the orbitals used in the definition of the DMET effective Hamiltonian are all localized and
we can construct larger active spaces  beyond the minimum model.
In Figure \ref{fig:larger_as_dmet} we see that adding purely virtual orbitals with d character
to create a (5e,10o) active space has only a modest impact on DMET's excitation energies.
In contrast, the addition of occupied orbitals related to the bonding between the iron and neighboring nitrogen atoms (yielding the (15e,10o) space),  leads to a more
significant lowering of the energy by multiple tenths of an eV, at both the CAS and NEVPT2 levels of theory.
When some or all the doubly occupied bonding-type orbitals and virtual d-type orbitals are present in the larger (11e,13o), (11e,16o), and (15e,15o) active spaces, the NEVPT2-DMET vertical excitation energy of the $\ce{^{4}E}$ state decreases to the range of 1.5 to 1.7 eV, significantly closer to the results of QDET and TDDFT.
Further discussion of the different active spaces and the orbitals included in them can be found in the Supporting Information.

\begin{figure}

\includegraphics[width=\columnwidth]{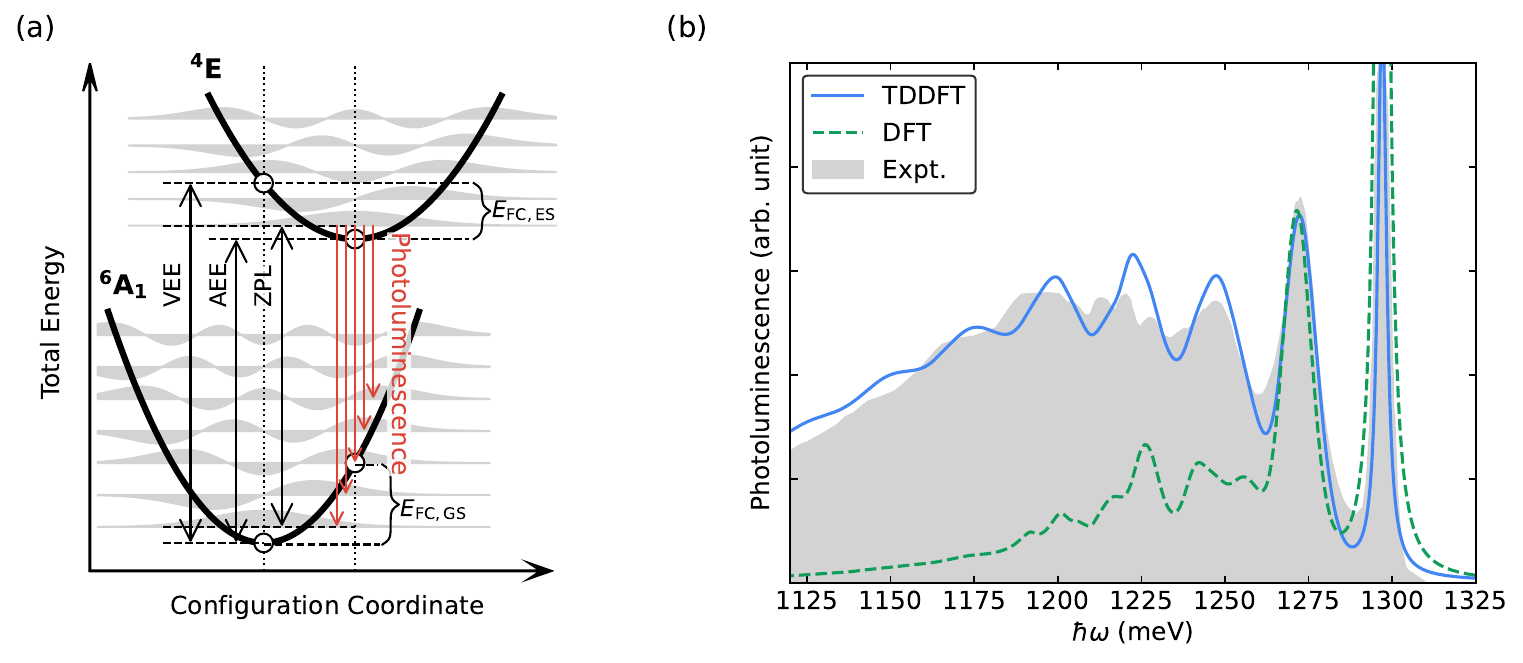}
\caption{ (a) Schematic representation of the relationship between the vertical excitation energy (VEE),
adiabatic excitation energy (AEE), zero-phonon line (ZPL), as well the Franck-Condon shifts. (b) Photoluminescence spectra for the $\ce{^{4}E} \to \ce{^{6}A_1}$ transition of the Fe impurity computed with both spin-flip TDDFT and spin-polarized DFT approaches using the HSE functional. The experimental photoluminescence spectrum from Ref.~\citenum{Baur1994} is measured at 2 K. The computed photoluminescence line shapes are horizontally shifted to align the position of the ZPL with the experimental data.}
\label{fig:vee_aee}

\end{figure}

To assess the accuracy of our converged QDET and DMET results, we turn to a comparison with experiments, for which, unfortunately, only a zero-phonon line (ZPL) measurement (1.30 eV\cite{Baur1994}) is available. To compare with ZPL data, we need to include the effect of excited-state geometric relaxation. Figure \ref{fig:vee_aee}(a) illustrates the relationships
among the vertical and adiabatic excitation energies and the Franck-Condon shifts.
Subtracting the excited-state Franck-Condon shift from the predictions of the vertical excitation energy of the different methods, we obtain an
estimate of adiabatic excitation energies. The latter are used to approximate the value of the ZPL, assuming that the vibrational zero-point energies in the excited and ground state potential energy surfaces cancel out.

To estimate the Franck-Condon shifts in the lowest $\ce{^{4}E}$ state, we performed excited-state geometry relaxations using spin-flip TDDFT, which enables the efficient computation of analytical forces acting on nuclei for highly correlated low-spin states, using the high-spin state as the reference.
Having considered the effects of supercell size and active space, we now focus on the largest supercell of 256 atoms and we derive adiabatic energies from vertical excitation energies obtained from DMET and QDET with (15e,15o) and (5e,5o) active spaces, respectively.
With TDDFT@PBE, we obtained an excited-state Franck-Condon shift of 0.11 eV, which is comparable to the values obtained with TDDFT and the HSE and DDH functionals, both yielding shifts of 0.08 eV.
From the results reported in Table \ref{tab:full_aee_data}, considering Franck-Condon shifts from TDDFT@PBE, we find that all
methods except CAS-DMET are accurate to within about 0.2 eV, with PBE-TDDFT and QDET producing a slight
underestimate of the ZPL energy, while NEVPT2-DMET and hybrid TDDFT give a small overestimate compared to the measured ZPL.
We note that  DMET requires an active space larger than QDET in order to obtain quantitatively accurate
excitation energies for the defect studies here. The reason is that in DMET the major portion of electron correlation is recovered in the CASSCF and post CASSCF step and thus an active space that allows for the representation of such correlation is needed. This requirement is well established in molecular CASSCF and post-CASSCF calculations.
The nature of the orbitals included in the active spaces of QDET and DMET differs and the two approaches capture electronic correlation in a different way.
As a result, a fixed active space size is not expected to yield results at the same level of accuracy for both methodologies. 
Further applications and comparisons of QDET and DMET in future may clarify the relative sizes of active spaces that each method requires for accurate results.

\begin{table}[H]

\caption{Adiabatic excitation energies (AEEs) of the lowest $\ce{^{4}E}$ state, computed with DMET, QDET, and TDDFT, as obtained by subtracting the excited-state Franck-Condon shift ($E_{\text{FC, ES}}$) calculated using spin-flip TDDFT with the PBE functional from the vertical excitation energies (VEEs) computed with the different methods in a 256-atom supercell. All quantities are in eV.}
\begin{tabular}{lccc}
Method & AEE & $E_{\text{FC,GS}}$ & $E_{\text{FC,ES}}$\\ \hline

PBE-TDDFT & 1.10 & 0.18 & 0.11 \\
HSE-TDDFT & 1.50 \\
DDH-TDDFT & 1.49 \\
CAS-DMET(15e,15o) & 1.94 \\
NEVPT2-DMET(15e,15o) & 1.46 \\
PBE-QDET (5e,5o) & 1.02 \\
PBE-QDET (51e,28o) & 1.03 \\
HSE-QDET (5e,5o) & 1.00 \\
HSE-QDET (47e,26o) & 1.02 \\
DDH-QDET (5e,5o) & 1.04 \\
DDH-QDET (43e,24o) & 1.06 \\
Exp ZPL\cite{Baur1994} & 1.30

\end{tabular}

\label{tab:full_aee_data}
\end{table}

\begin{figure}
    \centering
    \includegraphics[width=\columnwidth]{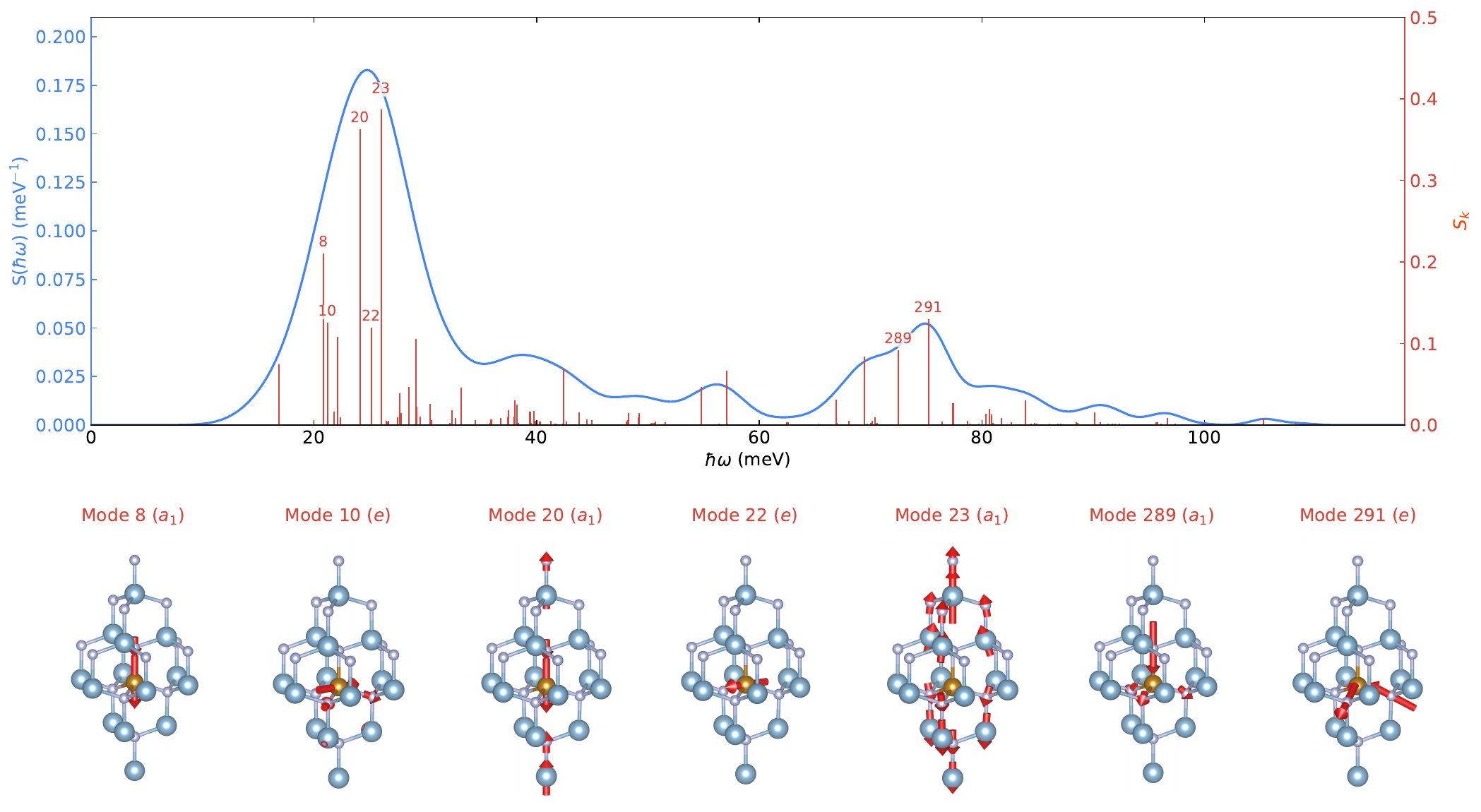}
    \caption{Partial Huang-Rhys factors ($S_k$, red lines) and the spectral density ($S(\hbar\omega)$, blue line), for the $\ce{^{4}E} \to \ce{^{6}A_1}$ transition calculated using spin-flip TDDFT with the HSE functional. The spectral density is defined as $S(\hbar\omega) = \sum_k S_k \delta(\hbar\omega - \hbar\omega_k)$. Vibrational modes that contribute significantly to the spectral density are indicated by red arrows on the atoms in the bottom panel, with their symmetries noted in parentheses.}
    \label{fig:v_modes}
\end{figure}

Using the relaxed excited-state atomic geometry of the lowest $\ce{^{4}E}$ state, we computed the vibrationally resolved photoluminescence spectrum for the $\ce{^{4}E} \to \ce{^{6}A_1}$ transition based on the Huang-Rhys theory~\cite{Alkauskas2014, Jin2021} and we compared it with experimental results~\cite{Baur1994}. As shown in Figure \ref{fig:vee_aee}(b), the photoluminescence line shape calculated using spin-flip TDDFT with the HSE functional is in close agreement with the experimental data. This strongly supports the assignment of the lowest excited state of the Fe impurity, from which the photoluminescence originates, as the $\ce{^{4}E}$ state, with its atomic geometry relaxation accurately described by spin-flip TDDFT. Compared to the ground-state geometry, the Fe-N bond in the axial direction shortens by 0.05 \AA\ in the excited-state geometry. Among the other three Fe-N bonds, two shorten by 0.025 \AA, while one elongates by 0.05 \AA, indicating a breakdown of the original $C_{3v}$ symmetry. This symmetry breaking arises from a coupling of the electronic states with both the symmetry-preserving $a_1$ and the symmetry-breaking $e$ phonon modes, as illustrated in Fig. \ref{fig:v_modes}. The figure shows the partial Huang-Rhys factors obtained by projecting the geometry displacements onto the vibrational modes and the spectral density $S(\hbar\omega)$. The quasi-local and local vibrational modes around the Fe impurity that contribute most significantly to $S(\hbar\omega)$ are also highlighted in Figure \ref{fig:v_modes}. Two major peaks can be identified in $S(\hbar\omega)$: one around 25 meV, originating from the vibration of the Fe atom resonating with the acoustic vibrational modes of AlN; another peak is found around 75 meV, originating from the vibration of the N atoms connected to the Fe atom. A comparison of $S(\hbar\omega)$ computed using PBE, DDH, and HSE functionals is provided in Figure S8 in the Supporting Information.

We also employed unrestricted spin-polarized DFT with the HSE functional to investigate the $\ce{^{4}E}$ excited state, constraining three more electrons in the spin-up channel than in the spin-down channel; this approach is referred to as the $\Delta$-SCF method in previous literature~\cite{Wickramaratne2019}. Although the computed adiabatic excitation energy is 1.44 eV, in satisfactory agreement with experiment, the photoluminescence spectrum computed using the excited-state geometry optimized with spin-polarized DFT significantly underestimates the peak intensities compared to the spin-flip TDDFT results and experiment (see Figure~\ref{fig:vee_aee}(b)). This discrepancy arises from the incorrect description of the strong static correlation effects for the $\ce{^{4}E}$ state by spin-polarized DFT. The single Slater determinant treatment of the $\ce{^{4}E}$ state in DFT neglects non-adiabatic coupling with higher-energy quartets, resulting in an underestimation of geometry relaxation. Specifically, the Fe-N bond in the axial direction shortens by only 0.03 \AA; among the other three Fe-N bonds, two shorten by less than 0.01 \AA, and one elongates by 0.015 \AA. This results in a deviation from the original $C_{3v}$ symmetry that is much less pronounced than that predicted by spin-flip TDDFT. Consequently, the Huang-Rhys factors and spectral densities are significantly smaller, as illustrated in Figure S9 in the Supporting Information.
This inaccurate treatment of the $\Delta$-SCF method also results in a spin expectation value ($\langle S^2 \rangle$) of 4.40, significantly different from the expected value of 3.75 for a spin quartet. In contrast, $\langle S^2 \rangle$ = 3.85 for the $\ce{^{4}E}$ state, when computed using spin-flip TDDFT, indicating a minor spin contamination. This observation suggests that spin-polarized DFT, or the $\Delta$-SCF approach, may not be suitable for studying low-spin excited states of spin defects with a high-spin ground state, even though they may yield reasonable results for excitation energies.


In summary, we have studied the excited states of the Fe impurity in AlN as a case study of transition metal impurities in insulators. We
compared the results of two quantum embedding techniques, QDET and DMET, and we studied the photoluminescence spectrum with spin-flip TDDFT.
The system studied here has proven to be challenging for other embedding
approaches, e.g. cRPA, and our study provides good agreement with experiments and a protocol to obtain converged results. An important aspect of our approach has been to leverage excited-state
geometry optimizations with spin-flip TDDFT to facilitate comparisons between
excitation energies from the embedding methods and experimental ZPL data.
We find that spin-flip TDDFT photoluminescence spectra are in good agreement with
experiment, in contrast to spin-polarized DFT.
Comparison to experiment is key to assessing the accuracy of the various
theoretical methods.
In this case study, we found that QDET and DMET provide satisfactory agreement with experiments with errors of the order of 0.1-0.2 eV.
Looking to the future, we anticipate that different quantum embedding methods, including QDET and DMET,
will be applied more widely to a variety of point defect systems in the search for useful qubits.
Careful comparisons between multiple techniques as we have pursued in
this work will enable the assessment of which embedding methods are most successful and for which classes of systems they are more adequate, thus facilitating accurate predictions of the properties of spin qubits.

\begin{acknowledgement}

L.O. was supported by the Chicago Center for Theoretical Chemistry Research Fellowship. The QDET and TDDFT calculations carried out by Y.J., V.W.-z.Y., and S.C. were supported by the Midwest Integrated Center for Computational Materials (MICCoM), as part of the Computational Materials Sciences Program funded by the U.S. Department of Energy (DOE), Office of Science, Basic Energy Sciences, Materials Sciences, and Engineering Division through Argonne National Laboratory. Part of the DMET calculations were supported by the U.S. DOE, Office of Science, National Quantum Information Science Research Centers and part were supported by the U.S. National Science Foundation QuBBE Quantum Leap Challenge Institute (NSF OMA-2121044). The computational resources were provided by the University of Chicago Research Computing Center's Midway cluster, the National Energy Research Scientific Computing Center, a DOE Office of Science User Facility supported by the Office of Science of the U.S. DOE under Contract No. DE-AC02-05CH11231, and the Argonne Leadership Computing Facility, a U.S. DOE Office of Science user facility at Argonne National Laboratory, which is supported by the Office of Science of the U.S. DOE under Contract No. DE-AC02-06CH11357.

\end{acknowledgement}

\section{Code Availability Statement}
The implementation of DMET is available within the pDMET code at \url{https://github.com/mitra054/pDMET}. 
The implementations of QDET and spin-flip TDDFT are available within the WEST code at \url{https://github.com/west-code-development/West}.
\begin{suppinfo}

DMET embedding subspace size and state-averaging, DMET orbitals in larger active spaces, detailed analysis of spectral densities and vibrational modes, comparison of spectral densities computed using HSE, DDH, and PBE functionals, comparison of spectral densities computed with spin-flip TDDFT and spin-polarized DFT

\end{suppinfo}

\bibliography{achemso-demo}

\end{document}


\tableofcontents
\addcontentsline{toc}{section}{Embedding Space Size in DMET}

\section{Embedding Space Size in DMET}

\begin{figure}

\includegraphics[width=\columnwidth]{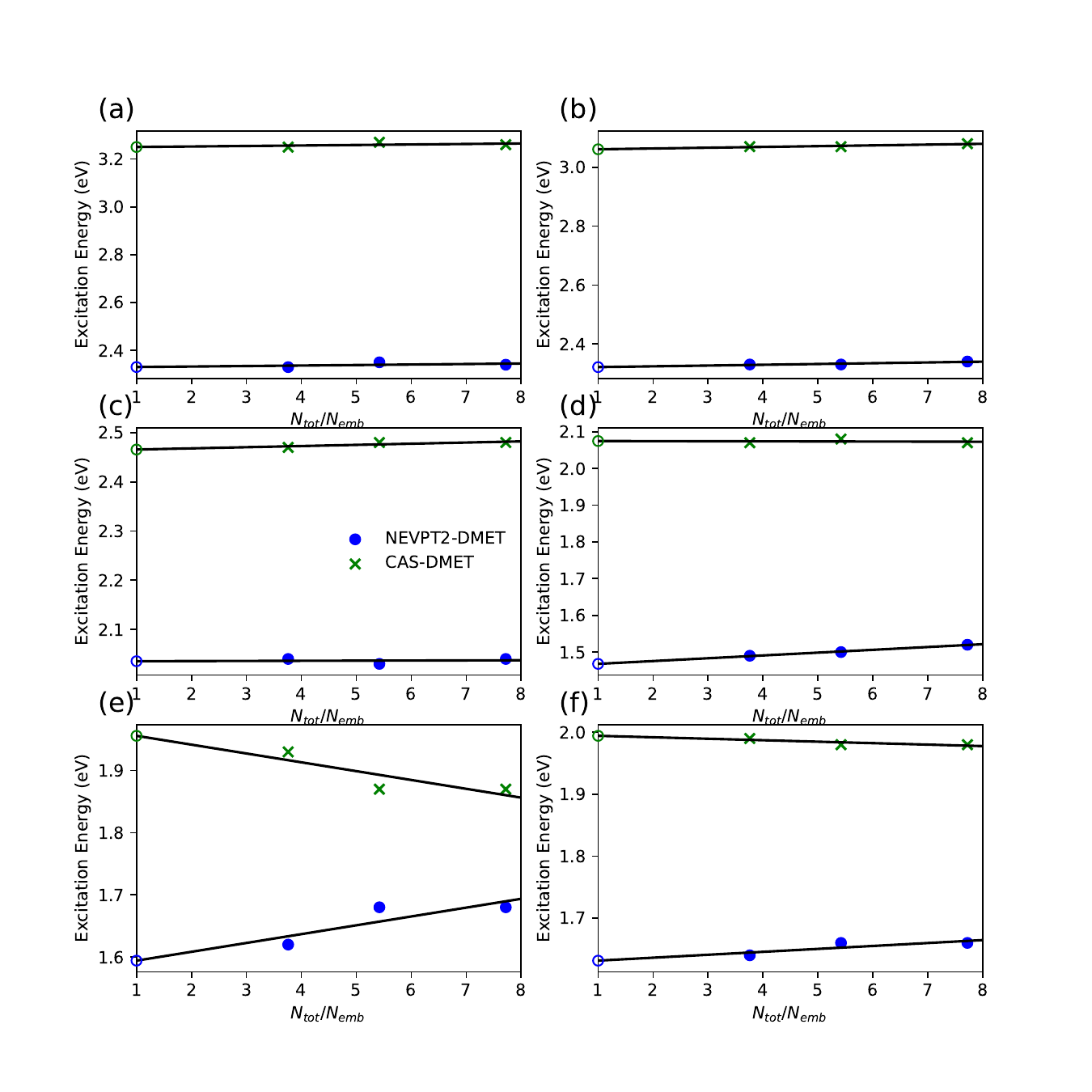}
\caption{NEVPT2-DMET and CAS-DMET vertical excitation energies of the lowest $\ce{^{4}E}$ excited 
state in a 108 atom supercell at different sizes of the embedding space for the following active spaces: (a) (5e, 5o), (b) (5e, 10o), (c) (15e,10o), (d) (11e,13o), (e) (11e,16o), and (f) (15e,15o). Open circles denote the extrapolated excitation energies in the nonembedding limit. The 
total number of basis functions, $N_{\text{tot}}$, is 1205.}
\label{fig:embed_size_dmet}

\end{figure}

\begin{figure}

\includegraphics[width=\columnwidth]{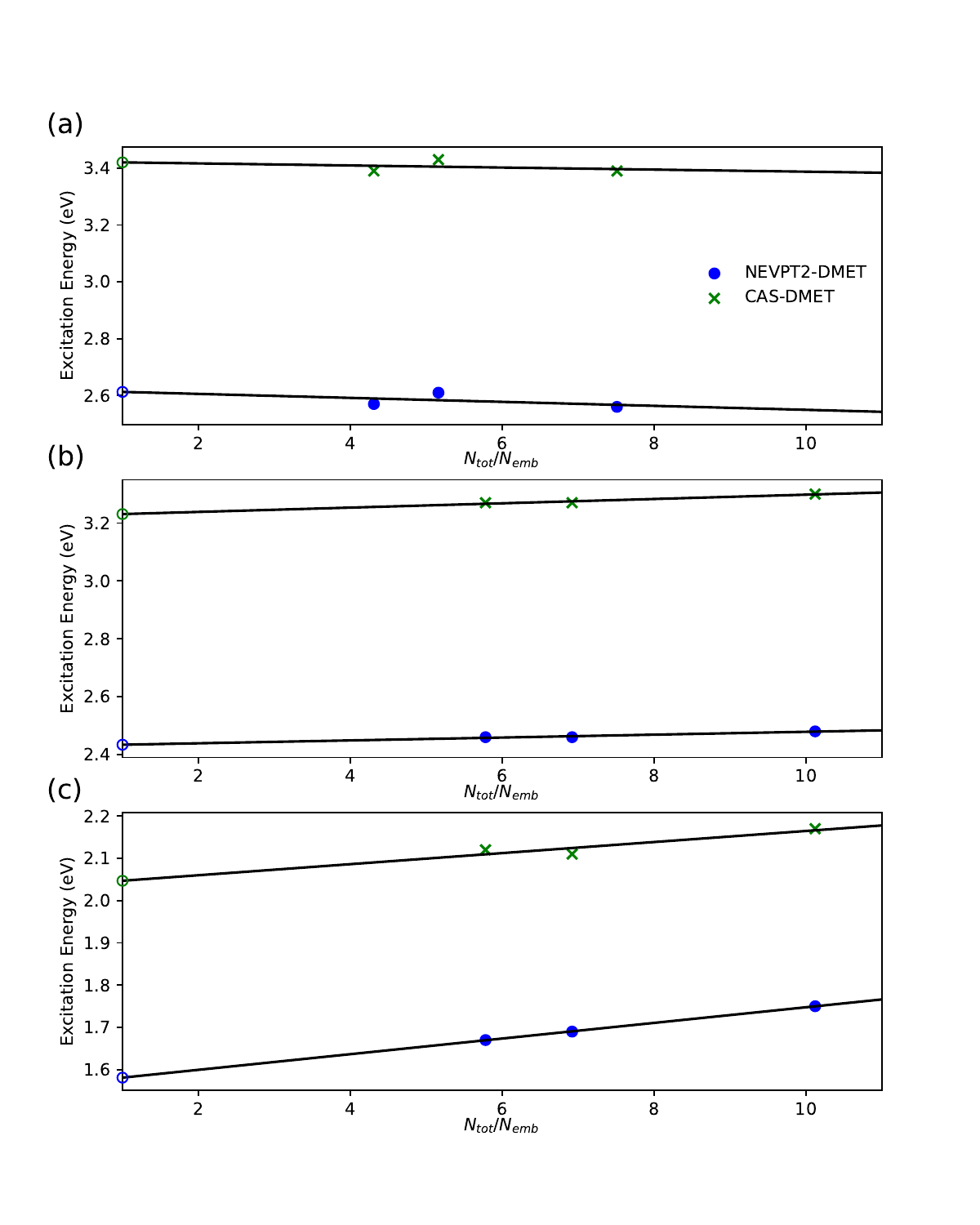}
\caption{NEVPT2-DMET and CAS-DMET vertical excitation energies of the lowest $\ce{^{4}E}$ excited 
state for (a) a 198 atom supercell with (5e,5o) active space, (b) a 256 atom supercell with (5e,5o) active space, and (c) a 256 atom supercell with (15e,15o) active space. Open circles denote the extrapolated excitation energies in the nonembedding limit. The 
total number of basis functions, $N_{\text{tot}}$, is 1194 for the 198 atom supercell and 1578 for the 256 atom supercell.}
\label{fig:large_cells_embed_size_dmet}

\end{figure}

One methodological choice in the practical use of DMET is the size of the 
embedding subspace.
Past applications of DMET to the nitrogen-vacancy center in 
diamond\cite{Haldar2023} and the oxygen vacancy in magnesium 
oxide\cite{Verma2023} have found that excitation energies are potentially 
sensitive to the embedding space size and employed linear extrapolations to the 
nonembedding limit.
For this case study, we find that DMET's excitation energy predictions for the 
Fe impurity are quite insensitive to the size of the embedding space.
As shown by the examples in Figures \ref{fig:embed_size_dmet} and \ref{fig:large_cells_embed_size_dmet}, the excitation energy 
predictions of both CAS-DMET and NEVPT2-DMET are only slightly changed by 
increasing the size of the embedding subspace and extrapolating to the 
nonembedding limit.
Across the different panels for different active spaces and supercells, we see that the 
extrapolated excitation energies typically differ by only a few hundredths of an eV 
from those obtained in the individual calculations and by no more than about a tenth of an eV in only a few cases.
In contrast, the prior DMET studies observed variations of multiple tenths 
of an eV or more in the excitation energy when extrapolating to the 
nonembedding limit.\cite{Haldar2023,Verma2023}
Wider application of DMET to more point defect systems in the future may 
clarify which cases exhibit substantial excitation energy changes under extrapolation and which do not.

\addcontentsline{toc}{section}{State Averaging in DMET}
\section{State Averaging in DMET}

Applications of DMET also require a choice of the number of states to be 
included in the state-averaged CASSCF portion of the calculation.
As noted in the main text, our results employ an average over 3 states: the 
$\ce{^{6}A_1}$ ground state and doubly degenerate $\ce{^{4}E}$ state.
While this number of states is sufficient for our focus on the lowest excited state, DMET can in general employ state-averaging over more states when exploring more of the spectrum and excitation energies are 
potentially sensitive to this choice. 
Here, we show in Table \ref{tab:sa_vee_data} that DMET's excitation energy predictions for the lowest $\ce{^{4}E}$ state differ by only a few hundredths of an eV when averaging over only 3 states or over 7, including higher energy $\ce{^{4}A_{2}}$, $\ce{^{4}A_{1}}$ and $\ce{^{4}E}$ states.
As discussed in the main text, the use of larger active spaces has a far 
more significant effect on DMET's excitation energies.

\begin{table}[H]

\caption{Vertical excitation energies of the $\ce{^{4}E}$ state in eV for CAS-DMET and NEVPT2-DMET with different amounts of state-averaging for the 256 atom supercell. A (5e,5o) active space is used in all cases.}
\begin{tabular}{lc}
Method   & Excitation Energy\\ \hline

CAS-DMET 7-SA    &  3.32 \\        
NEVPT2-DMET 7-SA  &  2.50 \\
CAS-DMET 3-SA &  3.27\\
NEVPT2-DMET 3-SA & 2.46 \\

\end{tabular}

\label{tab:sa_vee_data}
\end{table}

\addcontentsline{toc}{section}{Larger DMET Active Spaces}
\section{Larger DMET Active Spaces}

\begin{figure}[H]

\includegraphics[width=\columnwidth]{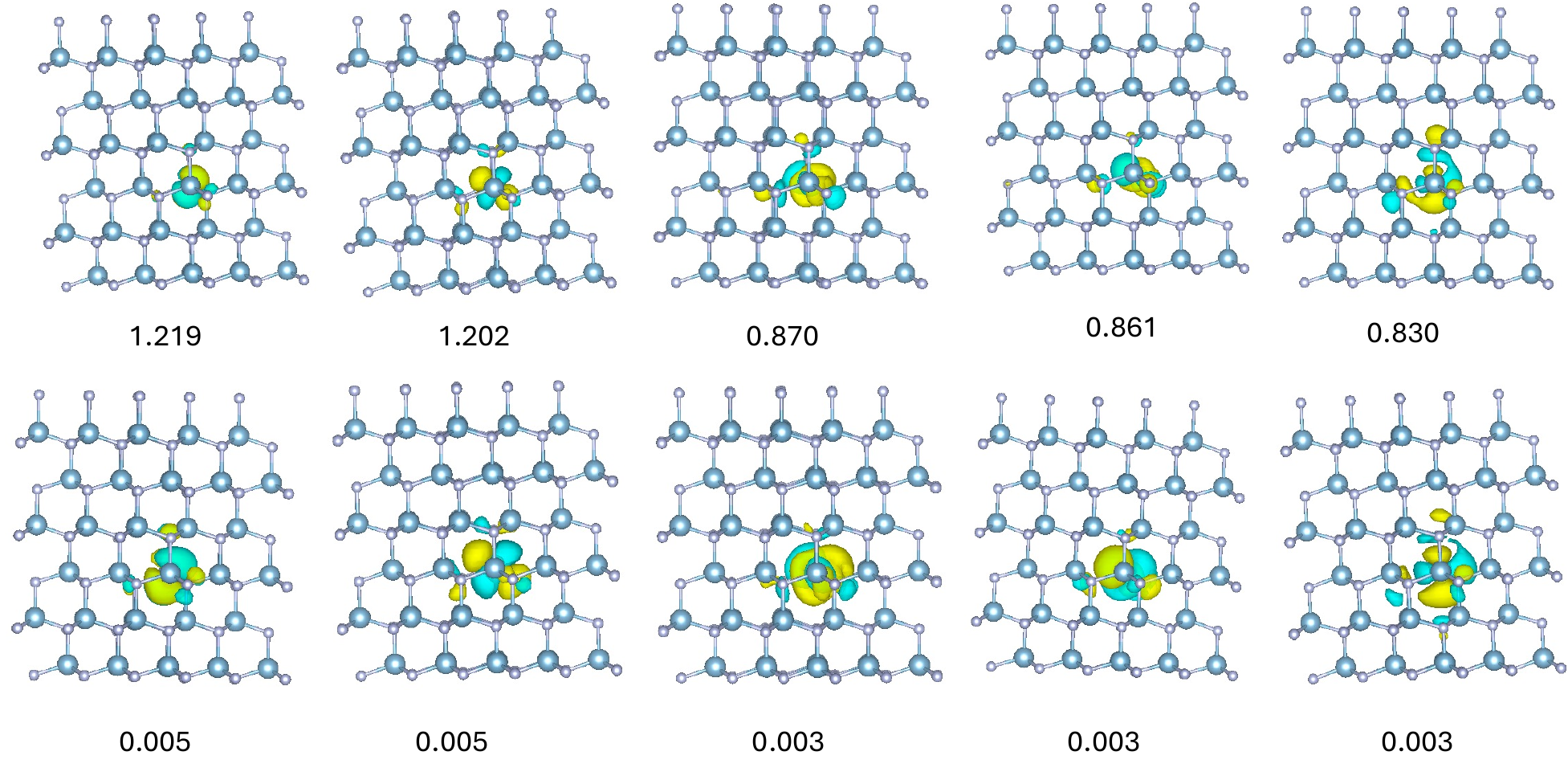}
\caption{Natural orbitals and occupation numbers for DMET (5e,10o) active space. }
\label{fig:5e10o_dmet_orbs}

\end{figure}

\begin{figure}[H]

\includegraphics[width=\columnwidth]{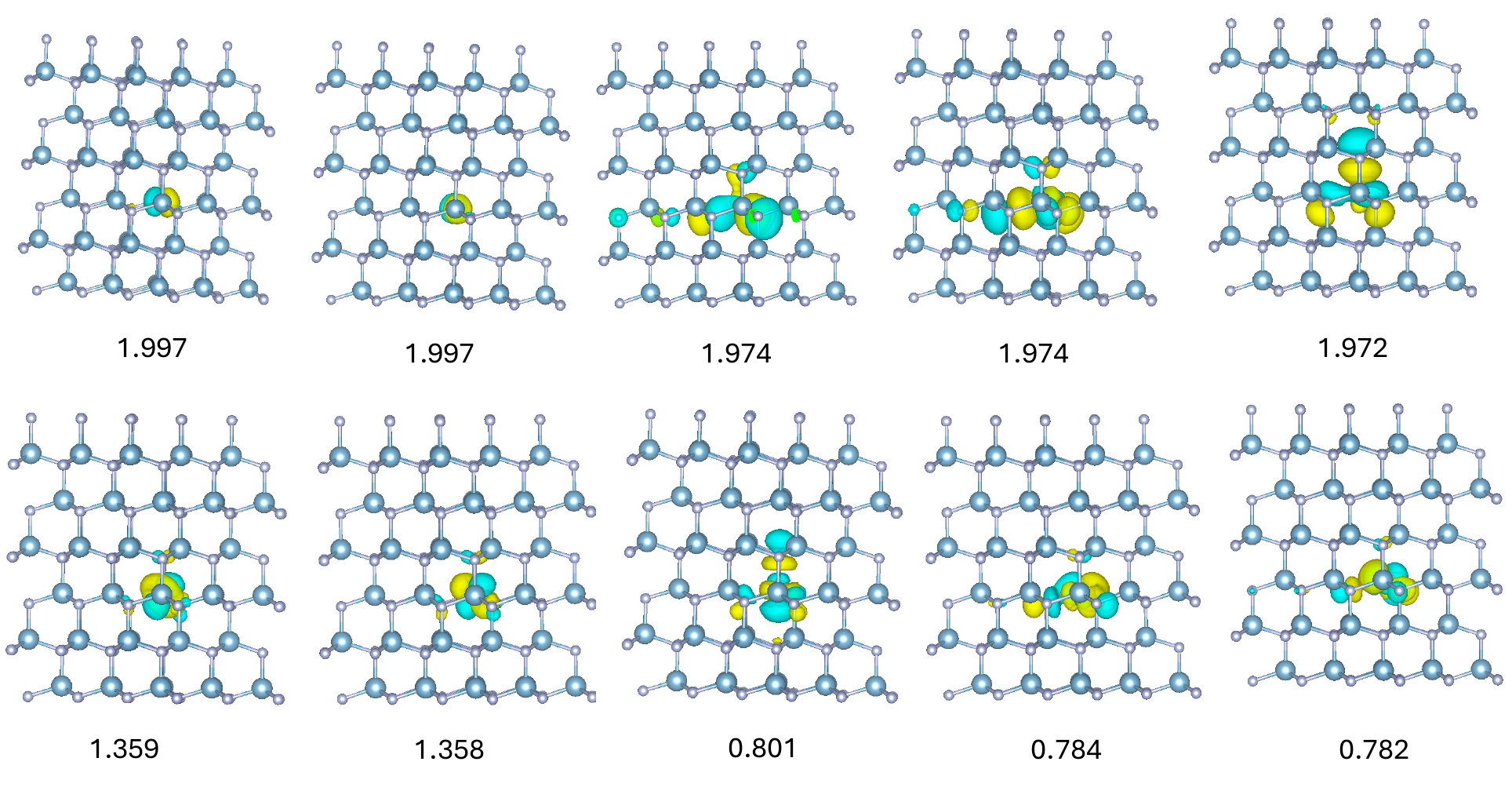}
\caption{Natural orbitals and occupation numbers for DMET (15e,10o) active space. }
\label{fig:15e10o_dmet_orbs}

\end{figure}

\begin{figure}[H]

\includegraphics[width=\columnwidth]{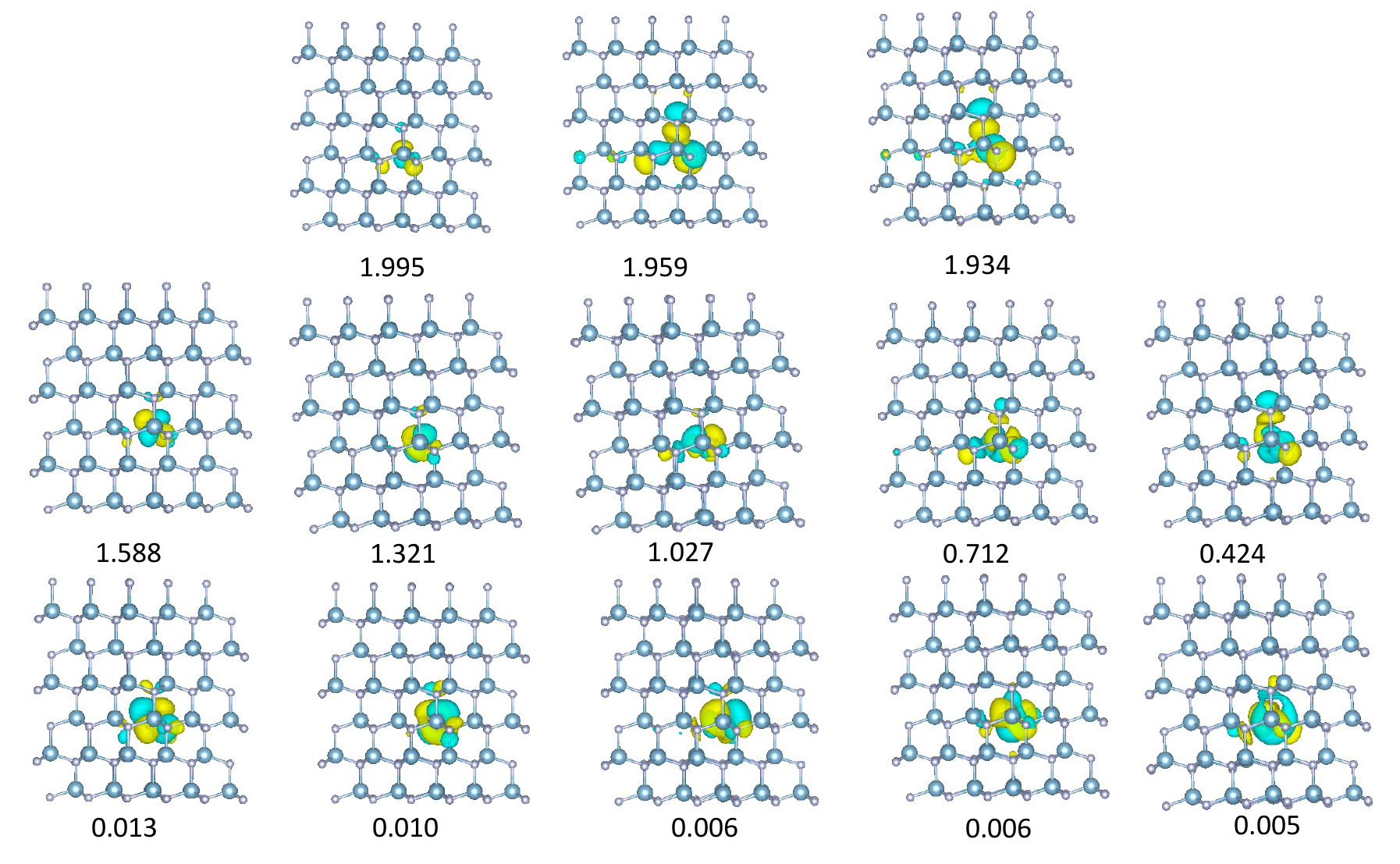}
\caption{Natural orbitals and occupation numbers for DMET (11e,13o) active space. }
\label{fig:11e13o_dmet_orbs}

\end{figure}

\begin{figure}[H]

\includegraphics[width=\columnwidth]{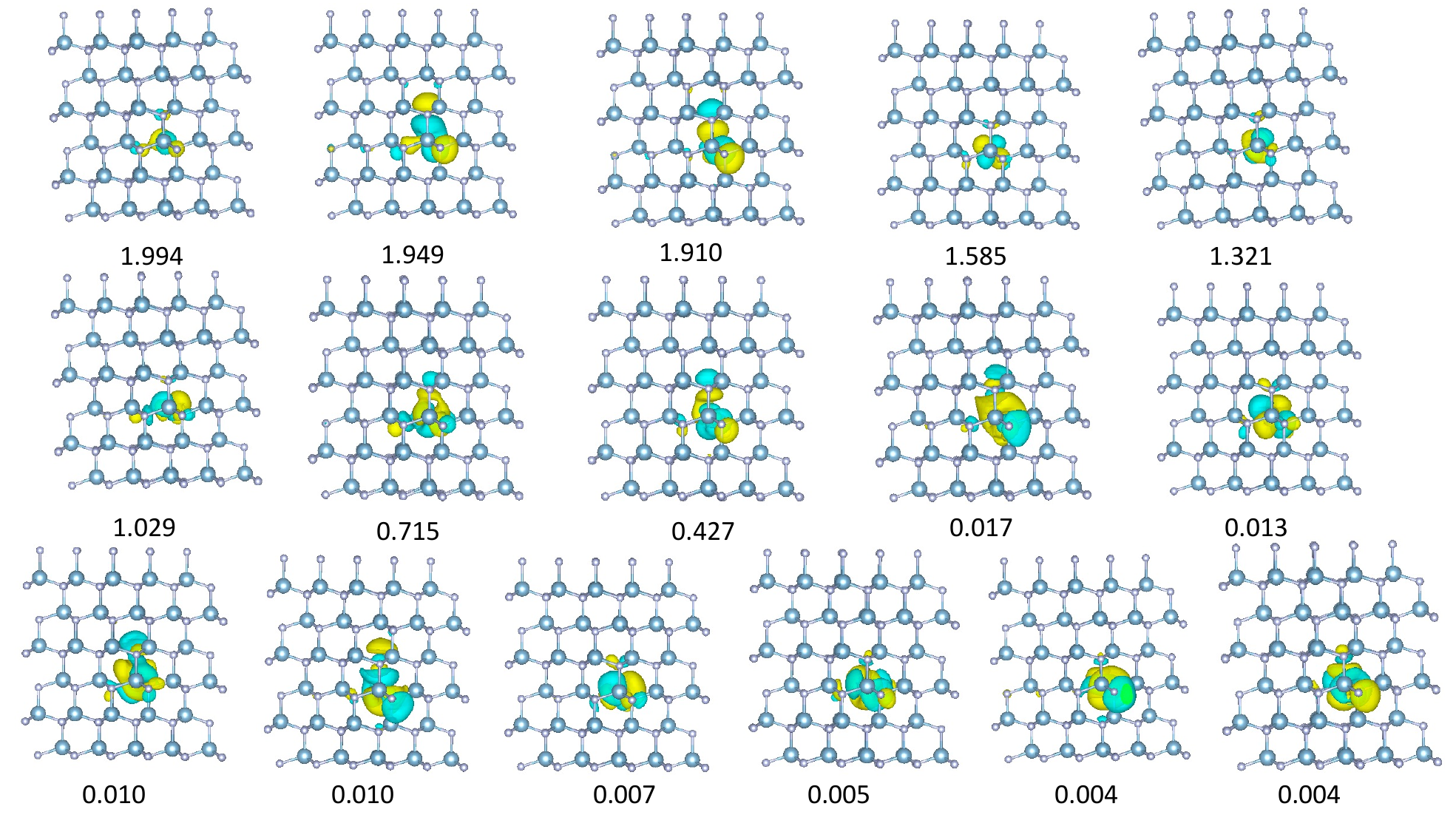}
\caption{Natural orbitals and occupation numbers for DMET (11e,16o) active space. }
\label{fig:11e16o_dmet_orbs}

\end{figure}

\begin{figure}[H]

\includegraphics[width=\columnwidth]{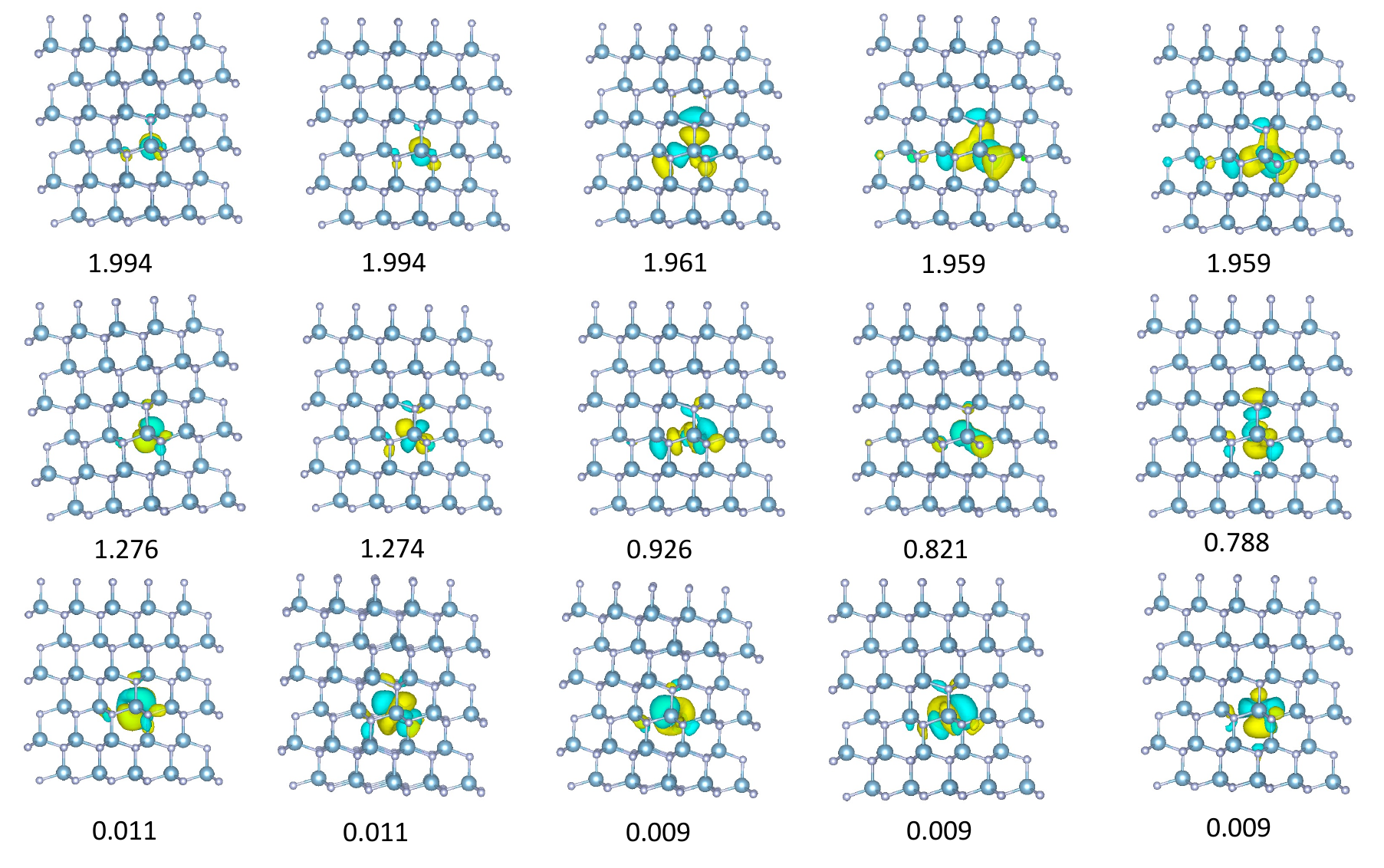}
\caption{Natural orbitals and occupation numbers for DMET (15e,15o) active space. }
\label{fig:15e15o_dmet_orbs}

\end{figure}

The orbitals for various enlarged DMET active spaces are shown in Figures \ref{fig:5e10o_dmet_orbs} through \ref{fig:15e15o_dmet_orbs} with their 
natural occupation numbers.
An initial attempt to go beyond the intuitive (5e,5o) space is to include 
the next set of virtual orbitals with d character as shown in Figure \ref{fig:5e10o_dmet_orbs}, but their occupation numbers remain close to zero and 
the impact on the DMET excitation energy is limited as shown in the main text.
The inclusion of occupied orbitals in the (15e,10o) and (15e,15o) active spaces yields two 
spectator orbitals of p character with occupations numbers close to 2 and 3 
orbitals related to bonding between the iron and neighboring nitrogens.
Attempts to use only the latter 3 orbitals and potentially a correlating trio of virtual orbitals with the respective (11e,13o) and (11e,16o) active space 
calculations resulted in the inclusion of one of the p character orbitals by CASSCF and 
yielded similar excitation energies to the (15e,15o) case.
The (15e,15o) active space is the largest that is feasible for our DMET implementation and provides an 
excitation energy in good agreement with other theoretical methods and 
experiment.

\addcontentsline{toc}{section}{Detailed Analysis of Spectral Densities and Vibrational Modes}
\section{Detailed Analysis of Spectral Densities and Vibrational Modes}

The partial Huang-Rhys factors, $S_k$, and the spectral density, $S(\hbar\omega) = \sum_k S_k \delta(\hbar\omega - \hbar\omega_k)$, for the photoluminescence of the $\ce{^{4}E} \to \ce{^{6}A}_1$ transition are shown in Figure 6 in the main text, along with several vibrational modes with the largest $S_k$. $S_k$ are computed as
\begin{equation}
    S_k = \dfrac{\omega_k \Delta Q_k^2}{2\hbar},
\end{equation}
where $\omega_k$ is the frequency of the $k$th vibrational mode, and
\begin{equation}
    \Delta Q_k = \sum_{\alpha = 1}^{N_{\mathrm{atoms}}} \sum_{i = x,y,z} \sqrt{M_\alpha} \left(R_{\ce{^{4}E}, \alpha i} - R_{\ce{^{6}A_1}, \alpha i} \right) e_{k, \alpha i}
\end{equation}
are the displacements between the equilibrium atomic geometry of the $\ce{^{6}A}_1$ state ($\mathbf{R}_{\ce{^{6}A_1}}$) and the $\ce{^{4}E}$ state ($\mathbf{R}_{\ce{^{4}E}}$), weighted by the square root of the atomic mass ($M_\alpha$), projected on the eigenvector of the $k$th vibrational mode ($\mathbf{e}_k$). Here, $\mathbf{R}_{\ce{^{4}E}}$ is obtained in the spin-filp TDDFT excited state geometry optimization calculation using the HSE functional. $\omega_k$ and $\mathbf{e}_k$ are obtained at the HSE level of theory using the frozen phonon approaches with the displaced geometries generated by the Phonopy code~\cite{phonopy-phono3py-JPCM, phonopy-phono3py-JPSJ}. The 192-atom supercell is used in the geometry optimization and the phonon calculations. $S_k$ and $S(\hbar\omega)$ are then used to compute the photoluminescence line shape based on the generating function approach~\cite{Alkauskas2014,Jin2021}.

Two major peaks can be identified in $S(\hbar\omega)$: one at around 25 meV, originating from the vibration of the Fe atom resonating with the acoustic vibrational modes of AlN; and another at around 75 meV, originating from the vibration of the N atoms connected to the Fe atom. Vibrational modes with both $a_1$ and $e$ symmetries contribute significantly to the spectral density, indicating symmetry breaking in the $\ce{^{4}E}$ state and suggesting the potential importance of non-adiabatic coupling and Jahn-Teller effects.

\addcontentsline{toc}{section}{Comparison of Spectral Densities Computed Using HSE, DDH, and PBE Functionals}
\section{Comparison of Spectral Densities Computed Using HSE, DDH, and PBE Functionals}

\begin{figure}
    \centering
    \includegraphics[width=12cm]{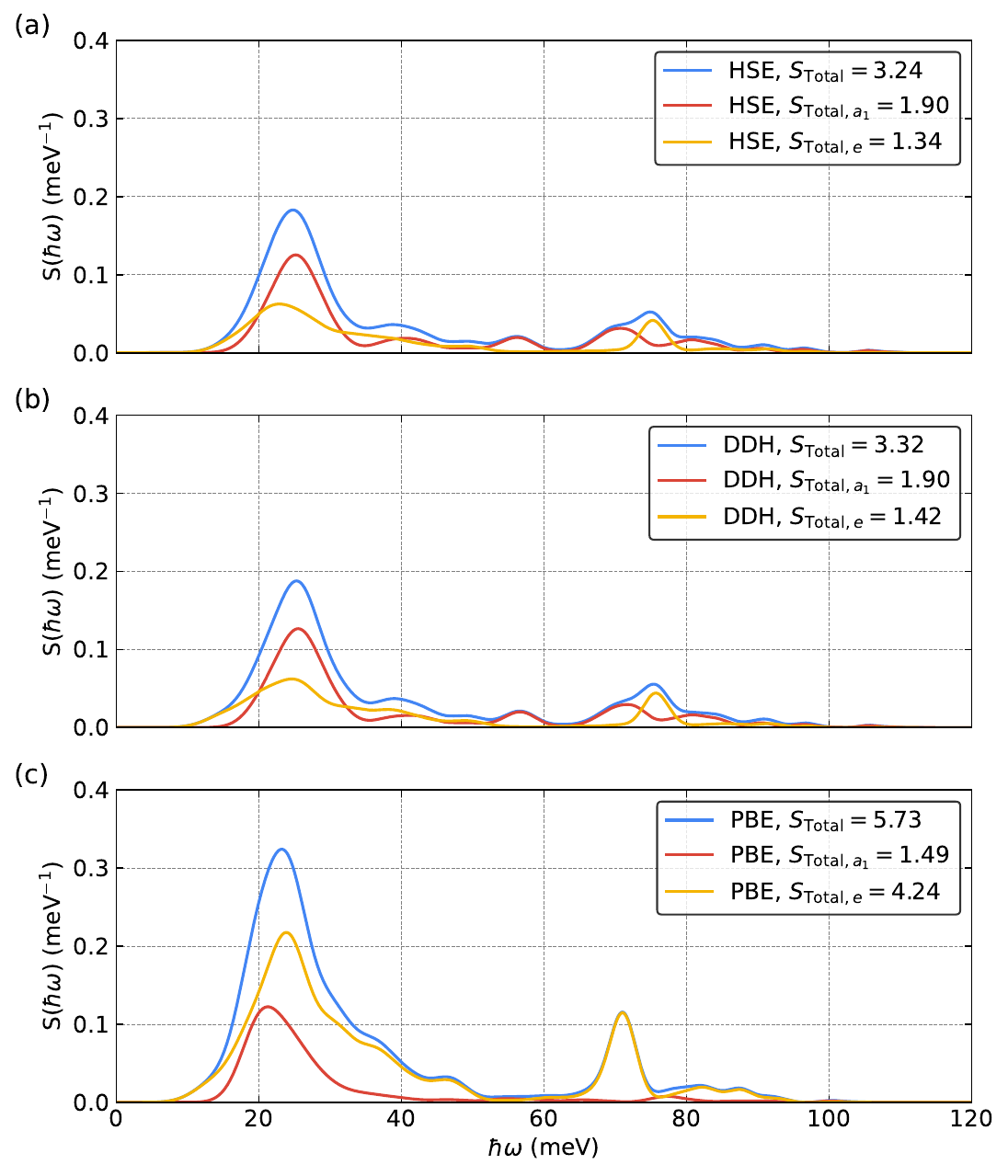}
    \caption{Spectral densities, $S(\hbar\omega)$, computed using spin-flip TDDFT with the (a) HSE, (b) DDH, and (c) PBE functionals. The spectral densities of the entire partial Huang-Rhys factors, partial Huang-Rhys factors of the $a_1$ symmetry, and partial Huang-Rhys factors of the $e$ symmetry are represented by blue, red, and yellow lines, respectively. The total Huang-Rhys factors are provided in the legend.}
    \label{fig:s_hse_vs_pbe}
\end{figure}

We computed the spectral densities of partial Huang-Rhys factors, $S(\hbar\omega)$, for the photoluminescence of the $\ce{^{4}E} \to \ce{^{6}A}_1$ transition, based on the equilibrium atomic geometries of the $\ce{^{4}E}$ state obtained using spin-flip TDDFT with both the HSE, DDH, and PBE functionals, as shown in Figure~\ref{fig:s_hse_vs_pbe}. Spectral densities obtained at the HSE and DDH levels of theory are very similar to each other. However, despite similar peak positions at around 25 meV and 70 meV, the intensity of the PBE $S(\hbar\omega)$ is significantly higher than that of the HSE $S(\hbar\omega)$, as reflected by the larger total Huang-Rhys factor of the former ($S_{\mathrm{Total}} = 5.73$) compared to the latter ($S_{\mathrm{Total}} = 3.24$). To understand the origin of this discrepancy, we computed the spectral densities of vibrational modes with $a_1$ and $e$ symmetries, $S_{a_1}(\hbar\omega)$ and $S_e (\hbar\omega)$, and displayed them in Figure~\ref{fig:s_hse_vs_pbe}. While $S_{a_1}(\hbar\omega)$ are comparable, $S_e(\hbar\omega)$ in the PBE case have a significantly higher intensity than in the HSE case. This is also reflected in the total Huang-Rhys factors: $S_{\mathrm{Total}, e} = 4.24$ for PBE, compared to $S_{\mathrm{Total}, e} = 1.34$ for HSE.

The difference in the contribution of the $e$ symmetry vibrational modes can be qualitatively understood by considering the non-adiabatic coupling between the lowest $\ce{^{4}E}$ state and the nearby quartet states through the symmetry breaking $e$ symmetry vibrational modes. The strength of the non-adiabatic coupling is related to the energy spacing between quartet states: smaller energy spacing results in stronger coupling. For a qualitative understanding, we computed the energy differences between the lowest $\ce{^{4}E}$ state and the $\ce{^{4}A_2}$, $\ce{^{4}E}$, and $\ce{^{4}A_1}$ states, which are the closest in energy, at the equilibrium geometries of the ground state. 
The average energy difference with the PBE functional is 0.072 eV, while with the HSE functional it is 0.126 eV. 
The inverse ratio of the average energy differences is $1 / (0.072 / 0.126) = 1.75$, which closely matches the square root of the ratio of $S_{\mathrm{Total}, e}$ at the PBE and HSE levels of theory, which is $\sqrt{4.24/1.34} = 1.78$. Although this qualitative analysis only considers the influence of energy differences on non-adiabatic coupling, it suggests that the overestimation of the spectral density with the PBE functional could be attributed to an underestimation of the energy spacing between quartets, which may result from the unsatisfactory treatment of exchange interactions in the semilocal functional.

\addcontentsline{toc}{section}{Comparison of Spectral Densities Computed with Spin-Flip TDDFT and Spin-Polarized DFT}
\section{Comparison of Spectral Densities Computed with Spin-Flip TDDFT and Spin-Polarized DFT}

\begin{figure}
    \centering
    \includegraphics[width=14cm]{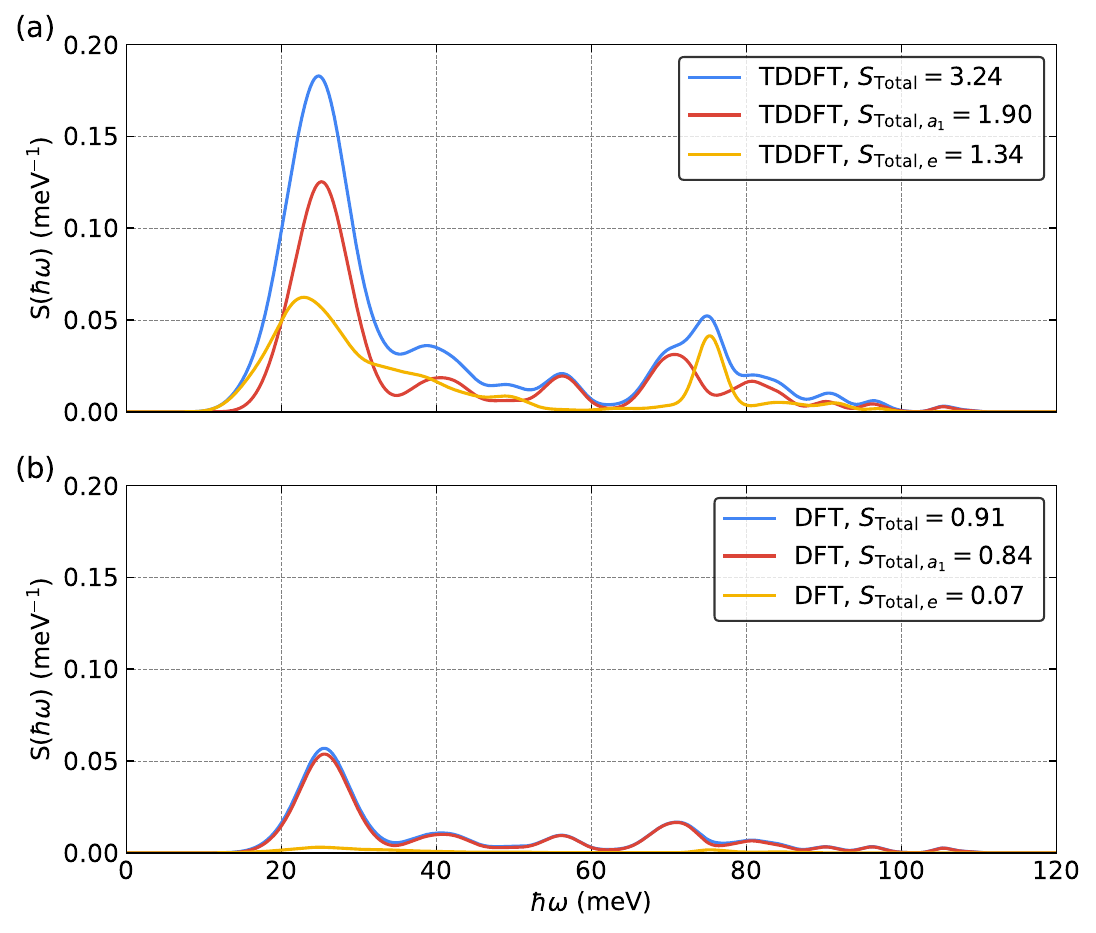}
    \caption{Spectral densities, $S(\hbar\omega)$, computed using (a) spin-flip TDDFT and (b) spin-polarized DFT with the HSE functional. The spectral densities of the entire partial Huang-Rhys factors, partial Huang-Rhys factors of the $a_1$ symmetry, and partial Huang-Rhys factors of the $e$ symmetry are represented by blue, red, and yellow lines, respectively. The total Huang-Rhys factors are provided in the legend.}
    \label{fig:s_tddft_vs_dft}
\end{figure}

We computed the spectral densities of partial Huang-Rhys factors, $S(\hbar\omega)$, for the photoluminescence of the $\ce{^{4}E} \to \ce{^{6}A}_1$ transition based on the equilibrium atomic geometries of the $\ce{^{4}E}$ state obtained using both spin-flip TDDFT and spin-polarized DFT with the HSE functional, as shown in Figure~\ref{fig:s_tddft_vs_dft}. The intensity of the spin-flip TDDFT $S(\hbar\omega)$ is significantly higher than that of the spin-polarized DFT $S(\hbar\omega)$, as reflected by the larger total Huang-Rhys factor of the former ($S_{\mathrm{Total}} = 3.24$) compared to the latter ($S_{\mathrm{Total}} = 0.91$). To understand the origin of this discrepancy, we computed the spectral densities of vibrational modes with $a_1$ and $e$ symmetries, $S_{a_1}(\hbar\omega)$ and $S_e (\hbar\omega)$, and displayed them in Figure~\ref{fig:s_tddft_vs_dft}. While $S_{a_1}(\hbar\omega)$ in the TDDFT case has a much higher intensity than in the DFT case, a significant discrepancy exists for $S_e(\hbar\omega)$: its intensity is comparable to $S_{a_1}(\hbar\omega)$ in the TDDFT case, while almost negligible in the DFT case. This is also reflected in the total Huang-Rhys factors: $S_{\mathrm{Total}, e} = 1.34$ for TDDFT, compared to $S_{\mathrm{Total}, e} = 0.07$ for DFT. The significant underestimation of $S_{\mathrm{Total}, e}$ in the spin-polarized DFT case can be attributed to the incorrect representation of the $\ce{^{4}E}$ state as a single Slater determinant, which neglects its non-adiabatic coupling with higher-energy quartet states, leading to an optimized geometry with negligible symmetry breaking. This emphasizes the importance of using an approach that allows for a multi-determinant description of the $\ce{^{4}E}$ state to accurately capture its geometric and optical properties.

\bibliography{achemso-demo}